\documentclass{iaus}
\usepackage{graphicx}

\usepackage{natbib}
\bibpunct{(}{)}{;}{a}{}{,} 

\usepackage{color}                             
\definecolor{change_col}{rgb}{1.0,0.0,0.0}     

\usepackage{url}

\newcommand{\ve}[1]{\mbox{\boldmath$#1$}}

\newcommand{\numb}[2]{{\left\{#1\right\}}_{#2}}
\newcommand{\unit}[2]{{\left[#1\right]}_{#2}}

\begin{document}

\title[Units of relativistic time scales and associated quantities]{Units of relativistic time scales and associated quantities}

\author[S.~Klioner \etal]{S.~Klioner$^1$,
N.~Capitaine$^2$,
W.~Folkner$^3$,
B.~Guinot$^2$,
T.-Y.~Huang$^4$,
S.~Kopeikin$^5$,
E.~Pitjeva$^6$,
P.K.~Seidelmann$^7$,
M.~Soffel$^1$}

\affiliation{
$^1$ Lohrmann Observatory, Dresden Technical University, 01062 Dresden, Germany\\
$^2$ 
SYRTE, Observatoire de Paris, CNRS, UPMC, 61 Av.~de l'Observatoire, 75014 Paris, France\\
$^3$ JPL m/s 301-150, 4800 Oak Grove Drive, Pasadena CA 91109, USA\\
$^4$ Nanjing University, Astronomy Department, 22 Hankou Road, 210093 Nanjing, China PR\\
$^5$ Department of Physics \& Astronomy, University of Missouri, Columbia, Missouri 65211, USA\\
$^6$ Institute of Applied Astronomy RAS, Nab Kutuzova 10, 191187 St Petersburg, Russia\\
$^7$ University of Virginia, 129 Fontana Ct, Charlottesville VA 22911-3531, USA}

\pubyear{2009}
\volume{261}  
\pagerange{1--4}
\setcounter{page}{1}
\jname{Relativity in Fundamental Astronomy:\\ Dynamics, Reference Frames, and Data Analysis}
\editors{S. Klioner, P. K. Seidelmann \& M. Soffel, eds.}

\maketitle

\begin{abstract}
  This note suggests nomenclature for dealing with the units of
  various astronomical quantities that are used with the relativistic
  time scales TT, TDB, TCB and TCG. It is suggested to avoid wordings
  like ``TDB units'' and ``TT units'' and avoid contrasting them to
  ``SI units''.  The quantities intended for use with TCG, TCB, TT or
  TDB should be called ``TCG-compatible'', ``TCB-compatible'',
  ``TT-compatible'' or ``TDB-compatible'', respectively. The names of
  the units second and meter for numerical values of all these
  quantities should be used with out any adjectives.  This suggestion
  comes from a special discussion forum created within IAU
  Commission~52 ``Relativity in Fundamental Astronomy''.
\end{abstract}

\maketitle

\section{Introduction}

In the current literature one can read different, sometimes
contradictory and illogical statements
about the units associated with
the values of various astronomical
parameters.  One sees wording like ``TDB units'' and ``TT units''
that is often contrasted to ``SI units''.
Such terminology is
often a source of confusion: no
serious discussion of how those ``TDB
units'' and ``TT units'' are defined can be found in
the literature.  The present note puts forward the
case for using clear and
consistent wording concerning the
units and values of various astronomical quantities to be used with
all standard astronomical reference systems (BCRS and GCRS) and time
scales (TT, TDB, TCB and TCG).  It is the result of a special
discussion forum created within IAU Commission~52 ``Relativity in
Fundamental Astronomy''.

\section{Quantities, values and units}
\label{section-quantities-values-and-units}

For the purposes of the present note, it is important
to distinguish clearly
between quantities and their numerical values.  According to \citet[][
definition 1.1]{VocabularyMetrology1993}, {\it quantity is an
attribute of a phenomenon, body or substance that may be
distinguished qualitatively and determined quantitatively}. {\it A
value (of a quantity)} is defined as {\it the magnitude of a
particular quantity generally expressed as a unit of measurement
multiplied by a number} \citep[][ definition
1.18]{VocabularyMetrology1993}.  The numerical values of quantities
are pure numbers that appear when quantities are expressed using some
units. For any quantity $A$ one has
\begin{equation}
\label{A=A[]A}
A=\numb{A}{}\, \unit{A}{},
\end{equation}
\noindent
where $\numb{A}{}$ is the numerical value (a pure number) of quantity
$A$ and $\unit{A}{}$ is the corresponding unit. Notations $\numb{A}{}$
and $\unit{A}{}$ for the numerical value and unit of a quantity $A$,
respectively, are recommended in ISO\,31-0 \citep{ISO31-0}.

The official definition of the concept of ``unit'' is given by
\citet[][ definition 1.7]{VocabularyMetrology1993}: {\it a unit (of
measurement) is a particular quantity, defined and adopted by
convention, with which other quantities of the same kind are
compared in order to express their magnitudes relative to that
quantity.}  Therefore, a unit is a sort of recipe of how an observer
can realize a specific physical quantity. The observer
can then express numerically all other quantities which have
the same physical dimensionality by comparing them
with that specific quantity  called ``unit''.

\section{SI second as the unit of proper time}

The official definition of the SI second can be found in
\citep{SI2006}:

\medskip

{\it

\noindent
The second is the duration of 9\,192\,631\,770 periods of the radiation
corresponding to the transition between the two hyperfine levels of the
ground state of the caesium 133 atom.

}

\medskip

\noindent
In the relativistic framework it is very important to realize that
this definition does not contain any hints on how the observer
realizing the second should move, or where (in what gravitational
field) that observer should be situated. From the relativistic point
of view this is the only correct approach to define a physical unit of
time.  A physical unit of time can be realized only by physical
observations.  One of the fundamental principles of General
Relativity, the Einstein Equivalence Principle, in combination with
the so-called locality hypothesis claims, in particular, that an
observer using only its proper time (the reading of an ideal clock
moving together with him) cannot judge how he is moving and how strong
the gravitation field along his trajectory is. Therefore, the concrete
second, called in the following the ``SI second'', can be realized by
any observer: a clock on the surface of the Earth, or on Mars or
on board a space vehicle far away from any planet. In this sense, the
SI second is the same for any observer and represents a recipe (as for
any unit of measurement: see
Section~\ref{section-quantities-values-and-units} above) to be
executed in order to realize unit time intervals locally. Thus, the SI
second is the unit of proper time and, as for proper time itself, it
can and should be realized only locally (but by any observer at an
arbitrary location and in an arbitrary gravitational field).  Hence,
it is clear that wording ``SI seconds on the geoid'' used in the
original definition of TAI does not mean that SI seconds are defined on the
geoid or can be realized only on the geoid. Such a wording actually
refers to the proper time of an observer on the geoid expressed in SI
seconds.

The official definition of the SI meter reads \cite{SI2006}:

\medskip

{\it

\noindent
The metre is the length of the path travelled by light in vacuum
during a time interval of 1/299\,792\,458 of a second.

}

\medskip

\noindent
The SI meter is,
therefore, based on the SI second and on the specific
defining value of the speed of light
$c=299\,792\,458$ m/s (assumed here to be
constant according to Special Relativity).


In the framework of General Relativity,
one should distinguish between
observable (or measurable) and coordinate quantities.  A measurable
quantity has dimension,
a unit, and gets a numerical value after
comparison with its unit. Its value is independent of the choice of
theory and reference systems.

A coordinate quantity has dimension, cannot be measured directly but
can get a numerical value after computation from observables with
proper theoretical (relativistic) relations. Its numerical value is
usually followed by
``second'', ``meter'' or
some combination according
to its dimension and the system of units used for the observables.
Its value depends on the choice of theory (General Relativity in
present IAU Resolutions) and reference systems.

In practice, all quantities not resulting directly from measurements
are coordinate quantities, such as time and space coordinates, orbital
elements, distances between remote points, and so on.

\section{Unit time intervals of different observers}

It is a common mistake to believe that intervals of proper time
$\Delta\tau_1$ and $\Delta\tau_2$ measured by different observers can
be ``uniquely'' and ``naturally'' compared to each other. The only way
to do so in General Relativity is to define a 4-dimensional
relativistic reference system having coordinate time $t$, establish a
relativistic procedure of coordinate synchronization of clocks with
respect to $t$, and convert the intervals of proper time
$\Delta\tau_1$ and $\Delta\tau_2$ of each
observer into corresponding
intervals of coordinate time $\Delta t_1$ and $\Delta t_2$. These two
intervals of coordinate time can
indeed be compared directly. If both
observers use SI seconds to measure their proper
times, and
$\numb{\Delta\tau_1}{}=1$ and $\numb{\Delta\tau_2}{}=1$ (i.e. both
proper time intervals have length of 1 SI second as realized by the
corresponding observer), the values of coordinate time intervals
$\numb{\Delta t_1}{}$ and $\numb{\Delta t_2}{}$ are in general
different.  It does not mean, however, that the observers use
different units of time. Only if 
the same units are used for
$\Delta\tau$ and $\Delta t$, numerical values $\numb{\Delta
t_1}{}$ and $\numb{\Delta t_2}{}$ are related to each other
according to the standard formulas of special- and
general-relativistic time dilations.

\section{Proper time, and coordinate times TCB and TCG}
\label{section-proper-time-TCB-TCG}

Along with the proper times of individual
observers, coordinate times are
indispensable for relativistic modelling of physical
processes. Coordinate times together with 3 spatial coordinates
constitute relativistic 4-dimensional reference systems. Full
definition of a reference system can be
achieved only by fixing its metric
tensor, as is done by the IAU \citep{IAU:2006,Rickman:2001} for the
standard reference systems BCRS and GCRS. Physical and mathematical
details of this definition can be found in \citet{SoffelEtAl2003}.

A relativistic
reference system can be associated with a set of rules
allowing one to label any phenomena or physical events with 4 real
numbers. One of these numbers is called coordinate time and the other
three are called spatial coordinates.  Any coordinate time is a
coordinate and, therefore, cannot be
measured directly. They can only
be {\it computed,} from the readings of
real clocks together with
additional parameters and information. For this computation one should
use the theoretical relation between the proper time of an observer and
coordinate time in General Relativity:
\begin{equation}
  \label{proper-time-vs-coordinate-time}
  {d\tau\over dt}={\left(-g_{00}(t,\ve{x}_{\rm obs}(t))
-{2\over c}\,g_{0i}(t,\ve{x}_{\rm obs}(t))\,{\dot x}^i_{\rm obs}(t)
-{1\over c^2}\,g_{ij}(t,\ve{x}_{\rm obs}(t))\,{\dot x}^i_{\rm obs}(t)\,{\dot x}^j_{\rm obs}(t)
\right)}^{1/2},
\end{equation}
\noindent
where $t$ is the coordinate time of a reference system having a metric
tensor with components $g_{00}$, $g_{0i}$ and $g_{ij}$ ($i$ and $j$
running from 1 to 3, and each component of the metric tensor being a function
of coordinate time $t$ and spatial position $\ve{x}$), and $\tau$ is the
proper time of an observer having position $\ve{x}_{\rm obs}(t)$ and
velocity ${\dot{\ve{x}}}_{\rm obs}(t)$ with respect to this reference
system. Einstein's implicit summation is used in the above formula.
In order to be useful this formula needs an initial condition of the
form
\begin{equation}
  \label{proper-time-initial-condition}
  \tau(t_0)=\tau_0,
\end{equation}
\noindent
where $t_0$ and $\tau_0$ are constants to be determined from the
procedure of clock synchronization (if there is only one observer
these two constants can be taken to be zero).  In the form written
above, Eq.~(\ref{proper-time-vs-coordinate-time}) allows one to
compute $\tau$ if $t$, $\ve{x}_{\rm obs}(t)$, ${\dot{\ve{x}}}_{\rm
  obs}(t)$ and the components of the metric tensor $g_{\alpha\beta}$
are given. This formula can be inverted (e.g.~numerically) in order
to compute $t$ for a given $\tau$.

We should note that Eq.~(\ref{proper-time-vs-coordinate-time}) is a
relation between quantities $\tau$ and $t$. By analogy with the rules
of quantity calculus, the values of $\tau$ and $t$ can be also related
by this formula if and only if the same units are used for both $\tau$
and $t$. This implies that if proper time $\tau$ is expressed in SI
seconds, then values of $t$ computed from numerical inversion of
Eq.~(\ref{proper-time-vs-coordinate-time}) should be also expressed in
SI seconds. A semantic difficulty here is that SI seconds can only be
realized for physically measurable proper time, whereas $t$ is a
non-measurable coordinate quantity related to the measurable $\tau$ by
Eqs.~(\ref{proper-time-vs-coordinate-time})--(\ref{proper-time-initial-condition}).
To accommodate this objection, one can agree to call the unit of time $t$
``SI-induced second''. These ``SI-induced seconds'' can be realized
only through the proper time of an observer.  In the following we will
call both ``SI seconds'' (used for proper times) and ``SI-induced
seconds'' (used for coordinate times) simply ``seconds''.

All the comments and arguments given above are equally correct
for both TCG and TCB.

Spatial coordinates $\ve{x}$ and $\ve{X}$ of BCRS and GCRS,
respectively, are also defined by the metric tensors of these
reference systems. The standard formulas of Special and General
Relativity assume that the locally measured
light speed in vacuum is equal to a constant quantity $c$ that
enters equations in many ways. In
practical calculations the specific value of $c$ from the definition
of the SI meter is always used. Therefore, if $t$ is expressed in
seconds, $x$ is expressed in meters.

Applying the same arguments of inheritance of the units to the
equations of motion for celestial bodies (Newtonian
equations of motion or
post-Newtonian EIH equations) we conclude that mass parameters
$\mu=GM$ of celestial bodies should also be expressed in the units of the SI.

We note that the views expressed above are closely related to the idea
of symbolic or abstract quantities and units developed in metrology
(see, \citet{deBoer:1994} and \citet{Emerson:2008}). In particular the
``SI-induced second'' discussed above appears as a symbolic second
\citep[see e.g.,][]{deBoer:1994}. A detailed discussion of these
concepts in the relativistic framework can be found in
\citet{Guinot:1997}.

\section{Scaled time scales TDB and TT}

The reasons for and the
mathematical details of the relativistic scaling of BCRS
and GCRS, and in particular their coordinate
times, are summarized by
\citet{Klioner:2008}. TT and TDB are conventional linear functions of
TCG and TCB, respectively. 
The definition of TT (given by IAU~2000
Resolution~B1.9 and IAU~1991 Resolution~A4)  can be written as

\begin{equation}
  \label{TT-definition}
  TT = F_G\,TCG
\end{equation}

\noindent
where $F_G=1-L_G$, and
$L_G=6.969290134\times10^{-10}$ is a defining
constant. TDB defined in IAU~2006 Resolution~3 is related to TCB as

\begin{equation}
  \label{TDB-definition}
  TDB = F_B\, TCB + TDB_0
\end{equation}

\noindent
where $F_B=1-L_B$, and
$L_B=1.550519768\times10^{-8}$ and $TDB_0$ are
defining constants.
Let us note here that TAI is a physical
realization of TT with a shift of $-$32.184~s for historical
reasons. Therefore, TAI is a realization of coordinate time
``TT$-$32.184~s''.
The difference between TAI and
an ideal realization of
``TT$-$32.184~s'' is only due to imperfections
of the participating clocks and the clock synchronization procedures.

The slopes $F_G$ and $F_B$ of both linear functions have the same
purpose: both TT and TDB should show no linear drift with respect to
proper times of observers
situated on the rotating geoid, i.e.~on the
surface of the Earth. Since this requirement
depends on our model of the
solar system (i.e.~on a planetary ephemeris and
on a number of astronomical and
geodetic constants), the latter requirement cannot be satisfied
exactly. Therefore, some conventional constants have been chosen in
the definitions of TT and TDB so that the requirement is satisfied
approximately, but with an accuracy totally sufficient for practical
purposes.
Similarly, the constant $TDB_0$ was chosen merely
to keep $TDB-TT$ approximately centered on zero.

Eqs.~(\ref{TT-definition}) and (\ref{TDB-definition}) define two new
quantities: coordinate time scales TT and TDB. As discussed in
\citet{Klioner:2008}, for practical reasons (keeping equations of
motion of celestial bodies and photons invariant)
these scalings of
coordinate time are accompanied by the corresponding scalings of
spatial coordinates $\ve{x}_{TDB}=F_B\,\ve{x}$ and
$\ve{X}_{TT}=F_G\,\ve{X}$ and mass parameters $\mu$ of celestial
bodies $\mu_{TDB}=F_B\,\mu$ and $\mu_{TT}=F_G\,\mu$.

The scaled coordinate times and spatial coordinates can be thought of
as defining two new reference systems: those with coordinates
$(TT,\ve{x}_{TT})$ and $(TDB,\ve{x}_{TDB})$.  These new reference
systems can be characterized by their own metric tensors, different
from those of the BCRS and GCRS.  Formulas $\mu_{TDB}=F_B\,\mu$ and
$\mu_{TT}=F_G\,\mu$ for the mass parameters follow from the
requirement to have the same form of the equations of motion with both
unscaled and scaled coordinates. This is an additional requirement
that does not immediately follow from the scaling of time and spatial coordinates.


In combination with (\ref{proper-time-vs-coordinate-time}),
Eqs.~(\ref{TT-definition}) and (\ref{TDB-definition}) define
how
TT and TDB are related to the proper time of any observer.
The proper time can be considered
a function of TT and TDB in a similar way to when we
considered it as
functions of TCG and TCB above. Therefore, the same
arguments as in Section~\ref{section-proper-time-TCB-TCG} can be used
to demonstrate that if proper times are expressed in SI seconds, both
TT and TDB are by inheritance expressed in SI-induced seconds or
simply in seconds.  Here ``by inheritance''
simply means that the
formulas linking TT and TDB to the other timescales, and ultimately to
proper times, provide a formal connection back to SI seconds.

Similarly, scaled spatial coordinates are expressed in SI-induced
meters or simply in meters. The same arguments allow us to conclude
that the scaled mass parameters are also
expressed in the units of the SI.

\section{Suggested terminology}

All these arguments allow us to suggest the following
nomenclature:

\begin{itemize}

\item[--] Avoid using
  the wording ``TDB units'' (``TDB seconds/meters''),
  ``TT units'' (``TT seconds/meters'') and avoid contrasting these
  terms with ``SI units'' (``SI seconds/meters'').

\item[--] All quantities intended for use with TDB should be called
  ``TDB-compatible quantities'' and corresponding values
  ``TDB-compatible values''.

\item[--] All quantities intended for use with TT should be called
  ``TT-compatible quantities'' and corresponding values
  ``TT-compatible values''.

\item[--] All quantities intended for use with TCB or TCG should be
  called ``TCB-compatible quantities'' or ``TCG-compatible
  quantities'' and the corresponding values ``TCB-compatible values'' or
  ``TCG-compatible values'', respectively. In the case of constants having
  the same value in BCRS and GCRS (e.g.~mass parameters $\mu=GM$ of
  celestial bodies) the value can be called ``unscaled''.
  Note
  that it is misleading to describe these values as ``SI-compatible''
  or ``in SI units'' since this does not distinguish unscaled values
  from TT- and TDB-compatible values. Such wording should be avoided.

\item[--] Consider that the numerical values of all
  above-mentioned quantities (TT-compatible, TCG-compatible,
  TDB-compatible and TCB-compatible) are expressed in the usual units
  of the SI. Avoid attaching any adjectives
  to the names of the units second and
  meter for numerical values of these quantities
\footnote{
  So, for example, it would be improper to say ``The interval is xx
  seconds of TDB''; the correct wording is ``The TDB interval is xx
  seconds''.
}
\footnote{
  An example of numerical values for a quantity is as follows: (i) the
  TCB/TCG-compatible value for $GM_E$ (the mass parameter of the Earth) is: 
$3.986004418\times10^{14}\ {\rm m}^3{\rm s}^{-2}$; 
(ii) the TT-compatible value for $GM_E$ is:
$3.986004415\times10^{14}\ {\rm m}^3{\rm s}^{-2}$; 
(iii) the TDB-compatible value for $GM_E$ is: 
$3.986004356\times10^{14}\ {\rm m}^3{\rm s}^{-2}$.
}
.

\end{itemize}

\begin{acknowledgement}
The authors gratefully acknowledge discussions with
George Kaplan, G\'erard Petit, and Patrick Wallace who also 
contributed to the text of this paper and to the proposed terminology.
\end{acknowledgement}

\end{document}